\documentstyle[epsfig]{elsart}

\begin{document}

\begin{frontmatter}

\title{
Spin alignment of $ K^*(892)^{\pm}$ mesons produced in neutron-carbon
interactions}

\collab{EXCHARM Collaboration}
\author[JINR]{A.N.~Aleev}
\author[HEPI]{N.S.~Amaglobeli}
\author[JINR]{V.P.~Balandin}
\author[IHEP]{A.P.~Bugorski}
\author[JINR]{C.V.~Cheshkov}
\author[JINR]{A.N.~Gaponenko}
\author[INRNE]{I.M.~Geshkov}
\author[IPG]{T.S.~Grigalashvili}
\author[JINR]{P.~Hristov\thanksref{mail}}
\author[JINR]{I.M.~Ivanchenko}
\author[JINR]{Z.M.~Ivanchenko}
\author[JINR]{M.N.~Kapishin}
\author[JINR]{N.N.~Karpenko}
\author[JINR]{V.D.~Kekelidze}
\author[JINR]{T.V.~Khristova}
\author[JINR]{Z.I.~Kojenkova}
\author[HEPI]{M.V.~Kopadze}
\author[JINR]{I.G.~Kosarev}
\author[JINR]{Yu.A.~Kretov}
\author[JINR]{A.S.~Kurilin}
\author[JINR]{N.A.~Kuz'min}
\author[HEPI]{R.A.~Kvatadze}
\author[JINR]{A.A.~Lavrentyev}
\author[AlmaAta]{A.A.~Loktionov}
\author[HEPI]{N.L.~Lomidze}
\author[JINR]{A.L.~Lyubimov}
\author[JINR]{D.T.~Madigojin}
\author[JINR]{A.S.~Mestvirishvili}
\author[JINR]{N.A.~Molokanova}
\author[JINR]{A.N.~Morozov}
\author[JINR]{A.A.~Panacik}
\author[Buk]{T.~Ponta}
\author[JINR]{Yu.K.~Potrebenikov}
\author[Buk]{T.~Preda}
\author[JINR]{A.G.~Skripnichuk}
\author[JINR]{L.A.~Slepets}
\author[JINR]{V.N.~Spaskov}
\author[JINR]{G.T.~Tatishvili}
\author[JINR]{A.L.~Tkatchev}
\author[JINR]{A.I.~Zinchenko}
\address[JINR]{Joint Institute for Nuclear Research, Dubna, Russia}
\address[IHEP]{Institute of High Energy Physics, Protvino, Russia}
\address[AlmaAta]{Institute of Physics and Technology,
Alma-Ata, Kazakhstan}
\address[Buk]{Institute of Atomic Physics,
Bucharest, Roumania}
\address[INRNE]{Institute for Nuclear Research and Nuclear Energy,
Sofia, Bulgaria}
\address[HEPI]{High Energy Physics Institute, Tbilisi State
University, Tbilisi, Georgia}
\address[IPG]{Institute of Physics, Georgian Academy of Sciences,
Tbilisi, Georgia}
\address[INPB]{Institute of Nuclear Problems of Belarus State
University, Minsk, Belarus}
\thanks[mail]{E-mail: hristov@sunse.jinr.ru, Peter.Hristov@cern.ch}

\begin{abstract}
A new precise measurements  of spin density matrix element $\rho_{00}$
of   $ K^*(892)^{\pm}$ mesons  produced  inclusively in neutron-carbon
interactions at $\sim  60$ GeV have been   carried out in  the EXCHARM
experiment  at the Serpukhov  accelerator.  The values of  $\rho_{00}$
obtained      in       the     transversity             frame      are
$0.424\pm0.011(stat)\pm0.018(sys)$    for        $K^*(892)^+$    and
$0.393\pm0.025(stat)\pm0.018(sys)$ for  $K^*(892)^-$.    Significant
$P_T$ dependence of $\rho_{00}$  has  been observed in $  K^*(892)^+$
production. 

The investigation  has been carried  out at the Laboratory of Particle
Physics, JINR.
\end{abstract}
\end{frontmatter}

 The inclusive hadroproduction of  $ K^*(892)^{\pm} $ has been studied
since more than 30 years, but the role of meson spin in the production
dynamics  is still  not understood  in    details. Spin phenomena   in
inclusive reactions with non-polarized  beam and target are  described
by spin density matrix   $\rho$ of  the   final state  particle.   The
$\rho_{00}$  element represents  relative intensity   of vector mesons
with zero $z$-component of the spin.  A  deviation of $\rho_{00}$ from
the value    $ \frac{1}{3}$ indicates spin     alignment.  A number of
phenomenological  models is  suggested   to predict spin behavior   in
vector-meson  inclusive  hadroproduction.     For   example,   in  the
semi-classical  model~\cite{DeGrand-Miettinen} the  spin alignment  is
expected for leading mesons, while for the non-leading ones $\rho_{00}
\sim \frac{1}{3}$ is anticipated.  An early experiment \cite{Blobel74}
reported  no spin  alignment  of $   K^*(892)^{\pm}$ produced  in $pp$
interactions at  12 and 24 GeV.  Significant  spin alignment  has been
observed   for leading vector mesons  inclusively  produced by charged
kaons: for   $K^*(892)^+$   and $K^*(892)^0$ in   $K^+p$   reaction at
8.2~GeV~\cite{Chliapnikov72},         32~GeV~\cite{Ajinenko80}     and
70~GeV~\cite{Barth83}, for   $K^*(892)^-$ and    $\bar{K}^*(892)^0$ in
$K^-p$    interactions       at         14.3~GeV~\cite{Paler75}    and
32~GeV~\cite{Arestov80}. 

The spin alignment of  $K^*(892)^{\pm}$ mesons produced inclusively in
neutron  beam has  been studied  for  the  first time  in the  EXCHARM
experiment and the results are presented in this paper . 

The EXCHARM  setup is a  forward magnetic spectrometer,  placed in the
neutral  beam 5N of Serpukhov accelerator.   Neutrons were produced on
the internal beryllium target by 70~GeV primary  protons at zero angle
to   the proton  orbit. A   20~cm    lead filter suppressed   $\gamma$
background. Accelerator magnets and a special sweeping magnet $SP-129$
rejected an  admixture of charged  particles in the beam.  The neutron
energy  spectrum \cite{spectrum} peaks at 58  GeV and has an effective
width  of 9   GeV.   The  $K^*(892)^\pm$ were   produced in    neutron
interactions with a 1.3g/cm$^2$ (1.5 cm) long carbon target located in
front of the  spectrometer.  The spectrometer analyzing  dipole magnet
SP-40A  causes a  transverse momentum   kick of 0.455  GeV/c with   an
alternative  polarity.     The   charged    particles   produced    in
neutron-carbon interactions  were detected by 11 proportional chambers
with 2  mm  wire spacing,  25 coordinate planes   in total (16  planes
upstream   and   9 --  downstream the     magnet).  Two  gas threshold
\u{C}herenkov counters filled with   freon and air at  the atmospheric
pressure were used for a charged  particle identification.  Hodoscopes
of two scintillator  counter  planes and three planes  of proportional
chambers were included in the trigger system.  The trigger requirement
selected events with at least 4  charged particles passing through the
spectrometer.   A  more detailed description  of  the apparatus can be
found elsewhere \cite{spectrometer}. 

The  analysis   has    been  performed using  $184.4     \times  10^6$
neutron-carbon interactions recorded in the experiment. 

The strange  resonances $K^*(892)^{\pm}$  have been selected  by their
decays into neutral kaon and charged pion: 
\begin{equation}
\begin{array}{cccccccc}
n & ^{12}C & \to & K^*(892)^{\pm}               &+ \quad X   &  &  &\\
  &        &     & ^{\mid} \hspace{-0.5em} \to  & K^0(\bar{K}^0)& \pi^{\pm} & &\\
  &        &     &                          & ^{\mid} \hspace{-0.5em}
  \to  & K_S& \to & \pi^+ \pi^- \\
\end{array}
\label{decay}
\end{equation}
The neutral
kaons have been identified via decays of short lived mode $K_S
\to \pi^+ \pi^-$. A pair of opposite charged particles ($V^0$) 
has been considered as a $K_S$ candidate if:
\begin{itemize}
\item{closest distance of approach between the $V^0$ tracks is less
than the experimental resolution (0.2~cm);}
\item{momentum ratio of the positive track to the negative one is less
than 5, in order to suppress the background from $\Lambda^0$ decays;}
\item{invariant mass of the $\pi^+ \pi^-$ system is within $\pm
30$~MeV$^2$ the PDG value of $K_S$;}
\item{\u{C}herenkov identification of each charged particle
is not consistent with kaon or proton/antiproton hypothesis.}
\end{itemize}
Each combination of a $K_S$ candidate with an additional track
has been regarded as $K^*(892)^{\pm}$ candidate if it meets
the following requirements:
\begin{itemize}
\item{ distance between the additional track and reconstructed
$K_S$ trajectory is less than 0.2~cm;}
\item{decay vertex of $K^*(892)^{\pm}$ is located within the target;}
\item{$K_S$ candidate life time is larger than $0.1\tau_S$ ($\tau_S$
is the PDG life time of $K_S$), in order to suppress the background
caused by interactions in the target;}
\item{\u{C}herenkov identification of additional charged particle
is not consistent with kaon or proton/antiproton hypothesis;}
\item{$K^*(892)^{\pm}$ candidate momentum is larger than 12~GeV/c.}
\end{itemize}

\begin{figure}[th]
\begin{picture}(140,140)
\epsfig{file=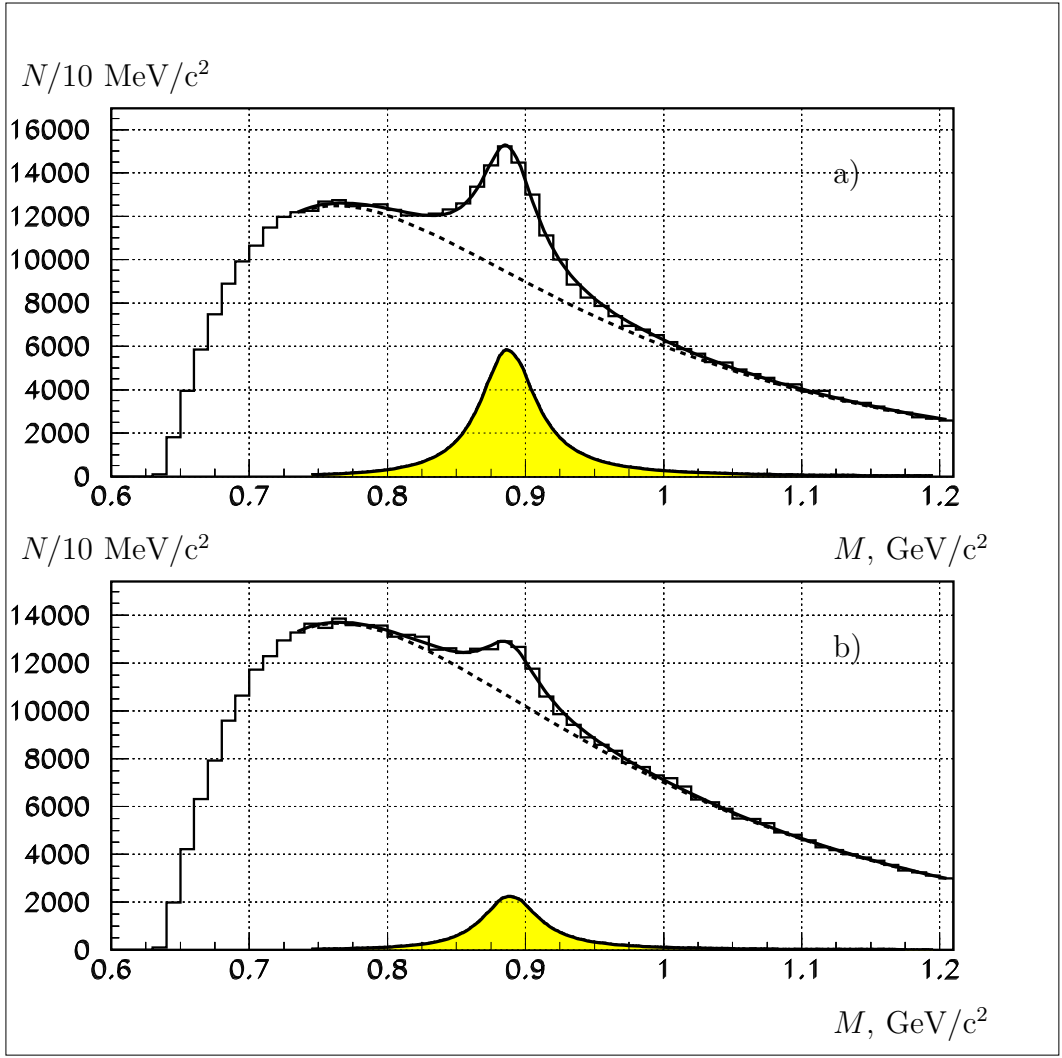,width=140mm}
\put(-30,3){$M$, GeV/c$^2$}
\put(-30,66){$M$, GeV/c$^2$}
\put(-138,66){$N/10$~MeV/c$^2$}
\put(-138,129){$N/10$~MeV/c$^2$}
\put(-30,53){b)}
\put(-30,116){a)}
\end{picture}
\caption{Invariant mass spectra of 
$K_S\; \pi^+$ (a) É $K_S\; \pi^-$ (b) combinations; 
the dashed curves represent the background; the shaded spectra show
the signal contribution.}
\label{fig:mass}
\end{figure}

The invariant mass ($M$)  spectra of selected  521480 $K_S \pi^+$  and
 553785     $K_S    \pi^-$    combinations    are      presented    in
 Fig.\ref{fig:mass}(a) and (b), respectively.  

Signals of decays (\ref{decay}) have been estimated as a result of the
$K_S\pi$-mass  spectrum   approximation  in the  region  of 0.74--1.20
GeV/c$^2$ by the  following expression representing a superposition of
background $BG(M)$ and signal $BW(M)$ with relative rate $a$: 
\begin{equation}
\frac{dN}{dM} = BG(M)[1+a BW(M)].
\end{equation}
The signal is   represented    by the relativistic   P-wave    ($l=1$)
Breit-Wigner  distribution    $BW(M)$ with   the mass-dependent  width
$\Gamma_R$: 
\begin{eqnarray}
BW(M) = \frac{M M_R \Gamma_R}{(M^2 - M_R^2)^2 + M_R^2 \Gamma_R^2},\\
\Gamma_R = \Gamma + 4\sigma^2, \quad \sigma^2 = \sigma^2_{res} +
\sigma^2_{bin},\nonumber \\
\Gamma = \Gamma_0 \left( \frac{p^*}{p_R^*} \right)^{2l+1} \: \frac{1+R
\, p^{*2}_R}{1+R\, p^{*2}},\nonumber 
\end{eqnarray}
where $p^*$ is the  pion momentum and  $p^*_R$ is the pion momentum in
the resonance maximum (both in the the resonance rest frame). The mass
resolution ($\sigma_{res}$) and  binning effects ($\sigma_{bin}$) have
been  considered   as simple modifications of    the  total width. The
$K^*(892)^\pm$ range parameter  $R \simeq 12.1$  is taken according to
its PDG value \cite{PDG}.   The background is  represented by a smooth
function 

\begin{equation}
BG(M) = b_1(M-M_{thr})^{b_2} \exp(b_3 M + b_4 M^2),
\end{equation}
where  $M_{thr}$ is the  resonance  mass threshold and $b_1,\dots,b_4$
are  free parameters.  The  influence of the  effective phase space on
the  shape  of resonance  has been  taken into  account in the product
$BG(M) \dot BW(M)$.  A total  of  $39180\pm 1070$ $K^*(892)^+$  decays
and $15280\pm 970$ $K^*(892)^-$ decays have  been observed (the errors
contain  also the   systematic  uncertainties   related  to the    fit
procedure). 

The evaluation of $\rho_{00}$ is based on the vector-meson decay
angular distribution 
\begin{equation}
W(|\cos\theta|) = \frac{3}{2} [1 - \rho_{00} + (3\rho_{00} - 1)
\cos^2\theta], 
\label{rho}
\end{equation}
where  the  angle   $\theta$   is  the    polar angle of     $\pi^\pm$
momentum-vector in the  decay $K^*(892)^{\pm} \to   K^0 \pi^\pm$.  The
measurements have been carried   out   in the transversity frame    of
$K^*(892)^{\pm}$ at rest.  The $z$-axis is  determined as a  normal to
the production plane, $y$-axis has  an opposite direction to the $K^*$
momentum defined in  the lab system.  The $x$-axis  is  defined by the
right-hand coordinate system. 

\begin{figure}[ht]
\begin{picture}(140,140)
\epsfig{file=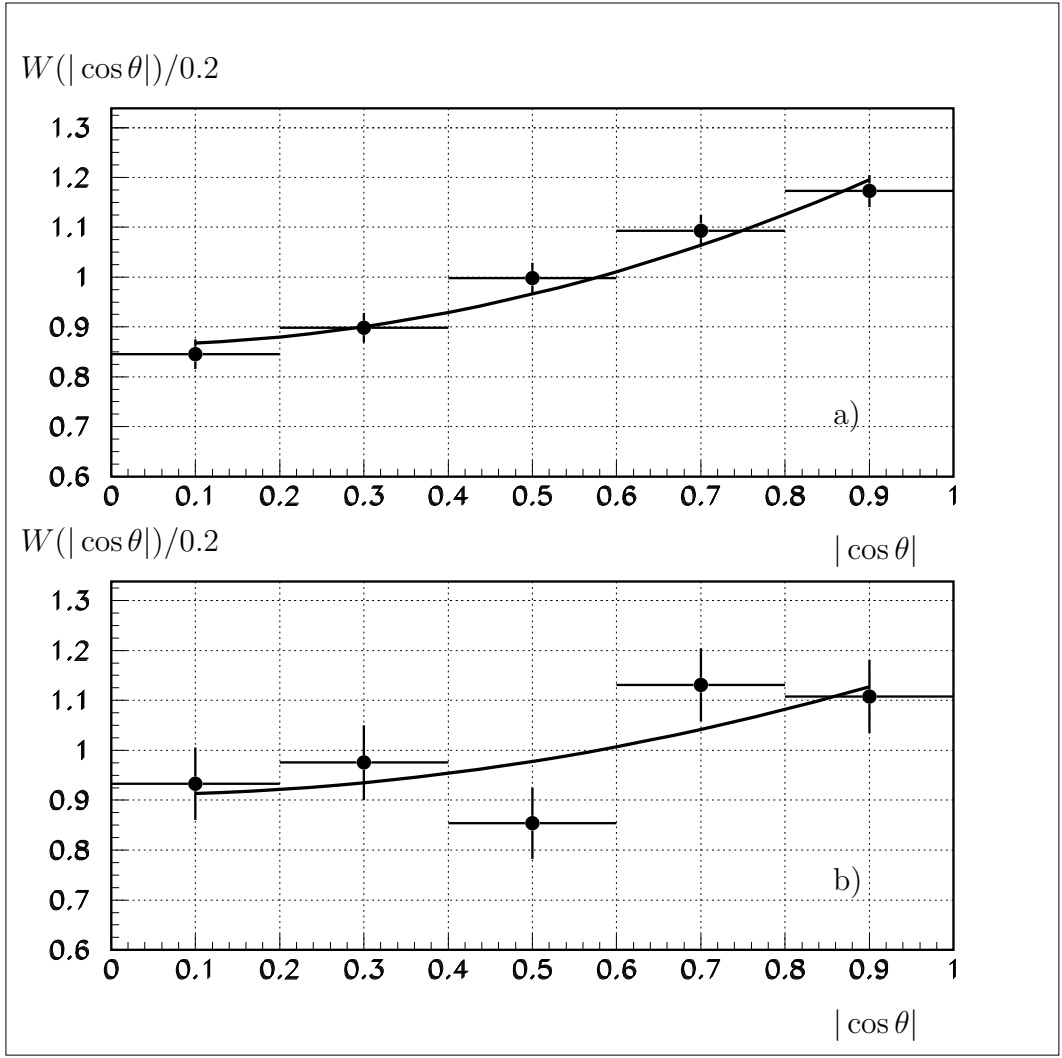,height=140mm,width=140mm}
\put(-30,3){$|\cos\theta|$}
\put(-30,66){$|\cos\theta|$}
\put(-138,67){$W(|\cos\theta|)/0.2$}
\put(-138,130){$W(|\cos\theta|)/0.2$}
\put(-30,22){b)}
\put(-30,84){a)}
\end{picture}
\caption{Angular distribution of the pion from the resonance decay
for $K^*(892)^+$ (a) and $K^*(892)^-$ (b) in the
transversity frame.}
\label{fig:rhot}
\end{figure}

To measure $\rho_{00}$ the  selected $K^*(892)^\pm$ have  been divided
in five intervals on  $|\cos\theta|$.  The relevant numbers of signal,
background   events, and corresponding   errors have been evaluated by
approximations of invariant mass spectra as described above in each of
the $|\cos\theta|$  intervals.   Supposing that the background  has no
spin  alignment, the corresponding  distribution   has been used  as a
measure   for   the acceptance    effects.    The  normalized  angular
distributions   of  pions from  $K^*(892)^+$  and  $K^*(892)^-$ decays
divided by the   normalized background   distributions for  acceptance
correction are given  in Fig.\ref{fig:rhot} (a) and (b), respectively,
together with the  approximation of $|\cos\theta|$-spectra by  formula
(\ref{rho})  (solid lines).   As a  result of  this approximation, the
$\rho_{00}$ have  been   obtained:  $\rho_{00}  =  0.424\pm0.011$  for
$K^*(892)^+$, and $\rho_{00} = 0.393\pm0.025$ for $K^*(892)^-$. 

The key assumption that the background has no spin alignment and could
be  used for    relative   acceptance corrections   has  been   proved
independently.   The  $K^*(892)^\pm$  inclusive   production  and  the
background   have   been    simulated  separately    by  FRITIOF model
\cite{FRITIOF} (with  no spin   alignment). Charged particle  tracking
through  the setup  was  realized in GEANT-based program~\cite{GEANT}.
The comparison  of $P_L, P_T^2$  and  charged multiplicity between the
simulated events  and the corresponding  experimental data has shown a
fair agreement, improved at  the  next stage by weighting   procedure.
Both simulations of the signal  and background give similar results on
the acceptance as  a function of  $|\cos\theta|$ which agree well with
the estimation from the experimental background. 

Systematic errors of $\rho_{00}$ have been  calculated by combining in
quadrature   the  contributions  from  detector   asymmetries  due to:
alternative polarity of magnetic field (0.01), simulated difference in
trigger  conditions related   to   the charged  multiplicity   (0.01),
different $K_S \pi^+$ and $K_S \; \pi^-$ acceptances (0.01) and varied
positions of internal target (0.002). As a result the systematic error
$\Delta=0.018$ of $\rho_{00}$  both for $K^*(892)^+$  and $K^*(892)^-$
has been obtained. 
 
\begin{figure}[th]
\begin{picture}(140,140)
\epsfig{file=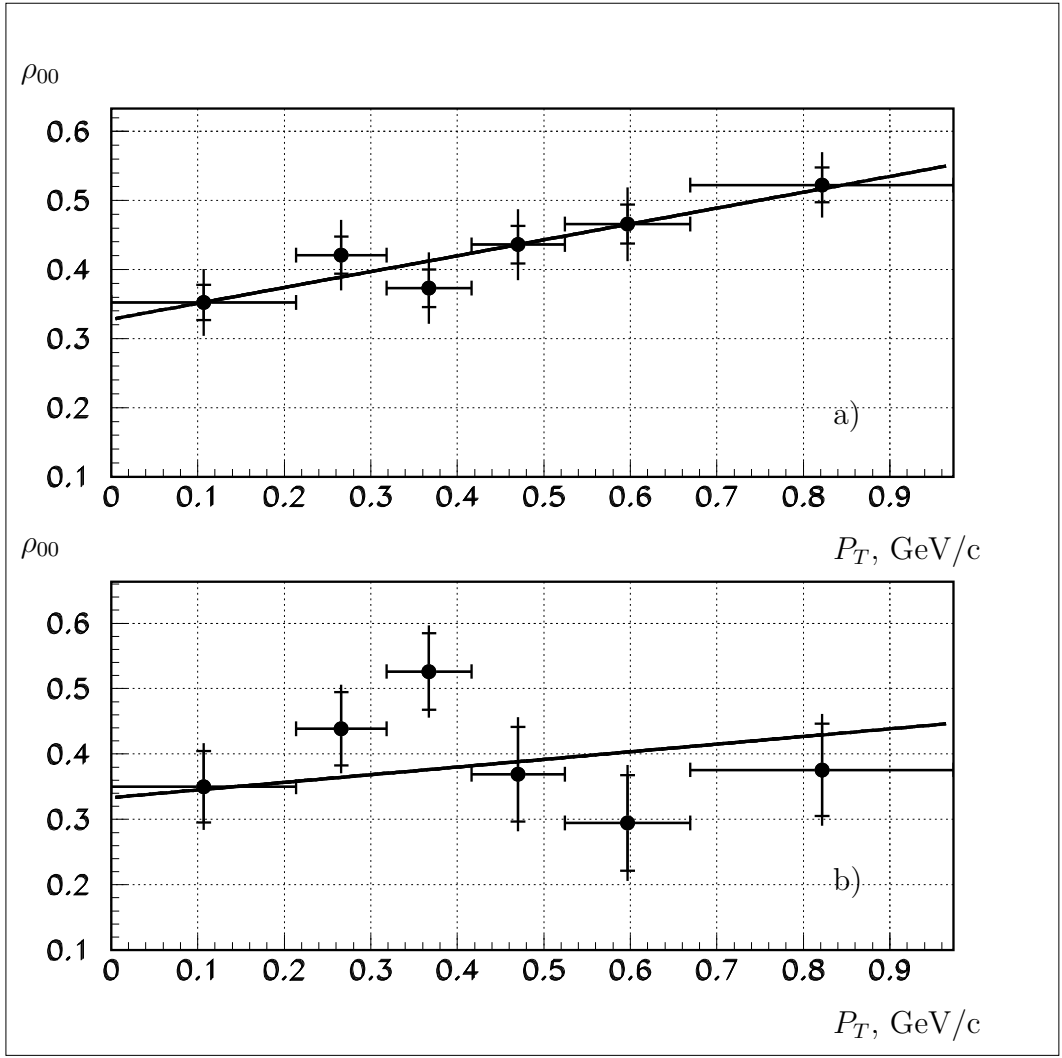,height=140mm,width=140mm}
\put(-30,3){$P_T$, GeV/c}
\put(-30,66){$P_T$, GeV/c}
\put(-138,67){$\rho_{00}$}
\put(-138,130){$\rho_{00}$}
\put(-30,22){b)}
\put(-30,84){a)}
\end{picture}
\caption{$P_T$-dependence of the spin density matrix element
$\rho_{00}$ for $K^*(892)^+$ (a) and $K^*(892)^-$ (b) 
in the transversity frame; the statistical errors are enlarged
quadratically by
the systematic ones.}
\label{fig:alivspt2}
\end{figure}

The  measurement  of  $\rho_{00}$  as a  function   of $P_T$ has  been
performed in     six   non-equidistant $P_T$   intervals\footnote{Each
interval on $P_T$ contains roughly  the same number of $K^*$ decays.}.
The same procedure as described above has been used for the estimation
of $\rho_{00}$  in each  of  $P_T$  intervals.   The systematic  error
$\Delta$ listed above has  been shared among  intervals ($i$) of $P_T$
according   to   the   formula      $\Delta_i=\sigma_i  \Delta   [\sum
(1/\sigma_i^2)]^{1/2}$, where $\sigma_i$   is   the statistical  error
\cite{PDG}.  The obtained $\rho_{00}$  dependences on $P_T$  are shown
in Fig.\ref{fig:alivspt2}. 
 One   can  see  a   clear $P_T$ dependence   of   the  matrix element
$\rho_{00}$ for the $K^*(892)^+$ mesons  while for $K^*(892)^-$ mesons
this   dependence is weaker.  The   $P_T$ dependence of $\rho_{00}$ is
approximated  by a linear function $\rho_{00}(P_T)  = a  + b P_T$. The
obtained  parameters for $K^*(892)^+$ are  $a = 0.328 \pm 0.023(stat)
\pm 0.037(sys)$ and $b = 0.23 \pm 0.045(stat) \pm 0.073(sys)$.  The
value of $a =  \rho_{00}(0)$ is  compatible  with 1/3, as expected  by
kinematic   reasons.  This indicates   the    absence  of  significant
additional   uncertainties   in  the   $K^*(892)^+$  analysis. If  the
constraint $a = \rho_{00}(0) = 1/3$  is fixed, the slope parameter
becomes $b = 0.22 \pm 0.022(stat) \pm 0.035(sys)$. 

Using the same constraint $a =  \rho_{00}(0) = 1/3$ in the $K*(892)^-$
case,   the slope $b  =   0.12 \pm  0.057(stat)  \pm 0.039(sys)$  is
obtained (Fig.\ref{fig:alivspt2} (b)) . 

{\bf  Conclusions}\\  Spin  density  matrix  element   $\rho_{00}$ for
leading vector mesons $K^*(892)^+$ has  been measured to be $\rho_{00}
= 0.424  \pm0.011(stat)\pm0.018(sys)$. This value  deviates from the
value  of $\frac{1}{3}$ indicating  the spin alignment of $K^*(892)^+$
produced     inclusively     in   neutron-carbon     interactions   at
$57\pm9$~GeV. It has been    obtained also that leading   vector meson
$K^*(892)^+$   has  spin alignment,   increasing   with  $P_T$.   Some
indications  on spin alignment of  non-leading  $K^*(892)^-$ have been
obtained, but with low  statistical  significance: $\rho_{00} =  0.393
\pm0.025(stat)\pm0.018(sys)$.  Thus    the   qualitative theoretical
expectations~\cite{DeGrand-Miettinen}   of leading particle effect  in
spin  alignment are confirmed.  The present precise  measurement is in
agreement with the result obtained in $K^+$ beam~\cite{Barth83}. 

The authors  are greatly   indebted to A.A.~Logunov,    N.E.~Tyurin, and
A.N.~Sis\-sa\-kyan  for their permanent  support   of present studies;   to
I.A.~Savin for his interest and valuable discussion. 

This work is supported  by the   Russian  Foundation for  Basic
Research, project 98-07-90294.

\end{document}